\documentclass[aps,prl,reprint,showpacs,superscriptaddress,floatfix,amsfonts,amsmath,amssymb]{revtex4-1}
\pdfoutput=1
\usepackage{graphicx}
\usepackage{subfig}
\usepackage{epstopdf}

\begin{document}

% Use the \preprint command to place your local institutional report
% number in the upper righthand corner of the title page in preprint mode.
% Multiple \preprint commands are allowed.
% Use the 'preprintnumbers' class option to override journal defaults
% to display numbers if necessary
%\preprint{}

%Title of paper
\title{Scheme for the detection of mixing processes in vacuum}
%\title{Vacuum mixing processes in a tightly focused laser field}
%\title{Enhancement of vacuum mixing processes in high numerical aperture from an annular beam configuration}

% repeat the \author .. \affiliation  etc. as needed
% \email, \thanks, \homepage, \altaffiliation all apply to the current
% author. Explanatory text should go in the []'s, actual e-mail
% address or url should go in the {}'s for \email and \homepage.
% Please use the appropriate macro foreach each type of information

% \affiliation command applies to all authors since the last
% \affiliation command. The \affiliation command should follow the
% other information
% \affiliation can be followed by \email, \homepage, \thanks as well.
\author{Fran\c{c}ois Fillion-Gourdeau}
\email{francois.fillion@emt.inrs.ca}
\affiliation{Universit\'{e} du Qu\'{e}bec, INRS-\'{E}nergie, Mat\'{e}riaux et T\'{e}l\'{e}communications, Varennes, Qu\'{e}bec, Canada J3X 1S2}

\author{Catherine Lefebvre}
\email{catherine.lefebvre@emt.inrs.ca}
\affiliation{Universit\'{e} du Qu\'{e}bec, INRS-\'{E}nergie, Mat\'{e}riaux et T\'{e}l\'{e}communications, Varennes, Qu\'{e}bec, Canada J3X 1S2}

%\author{Jean-Claude Kieffer}
%\email{kieffer@emt.inrs.ca}
%\affiliation{Universit\'{e} du Qu\'{e}bec, INRS-\'{E}nergie, Mat\'{e}riaux et T\'{e}l\'{e}communications, Varennes, Qu\'{e}bec, Canada J3X 1S2}

\author{Steve MacLean}
\email{steve.maclean@emt.inrs.ca}
\affiliation{Universit\'{e} du Qu\'{e}bec, INRS-\'{E}nergie, Mat\'{e}riaux et T\'{e}l\'{e}communications, Varennes, Qu\'{e}bec, Canada J3X 1S2}
\affiliation{Institute for Quantum Computing, University of Waterloo, Waterloo, Ontario, Canada, N2L 3G1}

%Collaboration name if desired (requires use of superscriptaddress
%option in \documentclass). \noaffiliation is required (may also be
%used with the \author command).
%\collaboration can be followed by \email, \homepage, \thanks as well.
%\collaboration{}
%\noaffiliation

\date{\today}

\begin{abstract}
A scheme for the detection of photons generated by vacuum mixing processes is proposed to observe the Quantum Electrodynamic photon-photon interaction. The strategy  consists in the utilization of a high numerical aperture parabolic mirror which tightly focuses two co-propagating laser beams with different frequencies. This produces a very high intensity region in the vicinity of the focus, where the photon-photon nonlinear interaction can then induce new electromagnetic radiation by wave mixing processes. These processes are investigated theoretically. The field at the focus is obtained from the Stratton-Chu vector diffraction theory, which can accommodate any configuration of an incoming laser beam. The number of photons generated is evaluated for an incident radially polarized beam. It is demonstrated that using this field configuration, vacuum mixing processes could be detected with envisaged laser technologies. 
\end{abstract}

% insert suggested PACS numbers in braces on next line
\pacs{42.50.Xa,12.20.-m,42.65.Ky}
% insert suggested keywords - APS authors don't need to do this
%\keywords{}

%\maketitle must follow title, authors, abstract, \pacs, and \keywords
\maketitle

%\section{Introduction}
%\label{sec:intro}

With recent advances in laser technology, that enable unprecedented intensity levels (above $10^{20}$ W/cm$^2$ \cite{RevModPhys.78.309}), there has been a surge of interest in the study and discovery of Quantum Electrodynamics (QED) processes in laser physics \cite{RevModPhys.84.1177}. One of the most important and surprising signatures of QED is the possibility of inducing wave mixing in the ``vacuum'', whereby electromagnetic radiation of novel frequencies are generated from the interaction between photons. This phenomenon is very similar to processes occurring in nonlinear media, such as four-wave mixing, harmonic generation and the propagation in birefringent material. These are the result of the nonlinear polarization and magnetization characterizing the medium response to an external electromagnetic field. According to the QED effective action \cite{dunne2005heisenberg}, an analogous feature also exists in vacuum owing to photon-photon interaction because these interactions also generate nonlinear polarization and magnetization \cite{rozanov1993four,Moulin1999137}. For relatively small field strengths obeying $|\mathbf{E}|,|\mathbf{B}| \ll E_{S}$ (where  the Schwinger field is $E_{S} = 1.3 \times 10^{18}$ V/m), the leading order expression in the fine coupling constant $\alpha \approx 1/137$ of the vacuum response scales like $\sim \alpha^{2}$ and is cubic in the electromagnetic field. As a consequence, the effect is very weak for field strengths reached by current lasers and thus, it has eluded experimental verification so far \cite{moulin2000}. Nevertheless, many observables have been studied and proposed theoretically to investigate the QED nonlinearity, taking advantage of the formal analogy with optics. These include four-wave mixing \cite{rozanov1993four,Moulin1999137}, vacuum birefringence \cite{Iacopini1979151,1998ftqp.conf...29D,Heinzl2006318,0034-4885-76-1-016401} and second harmonic generation \cite{PhysRevLett.63.2725}. All of these processes occur due to the fact that QED induces a cubic nonlinearity. 

\begin{figure}[h]
	\centering
		\includegraphics[width=0.40\textwidth]{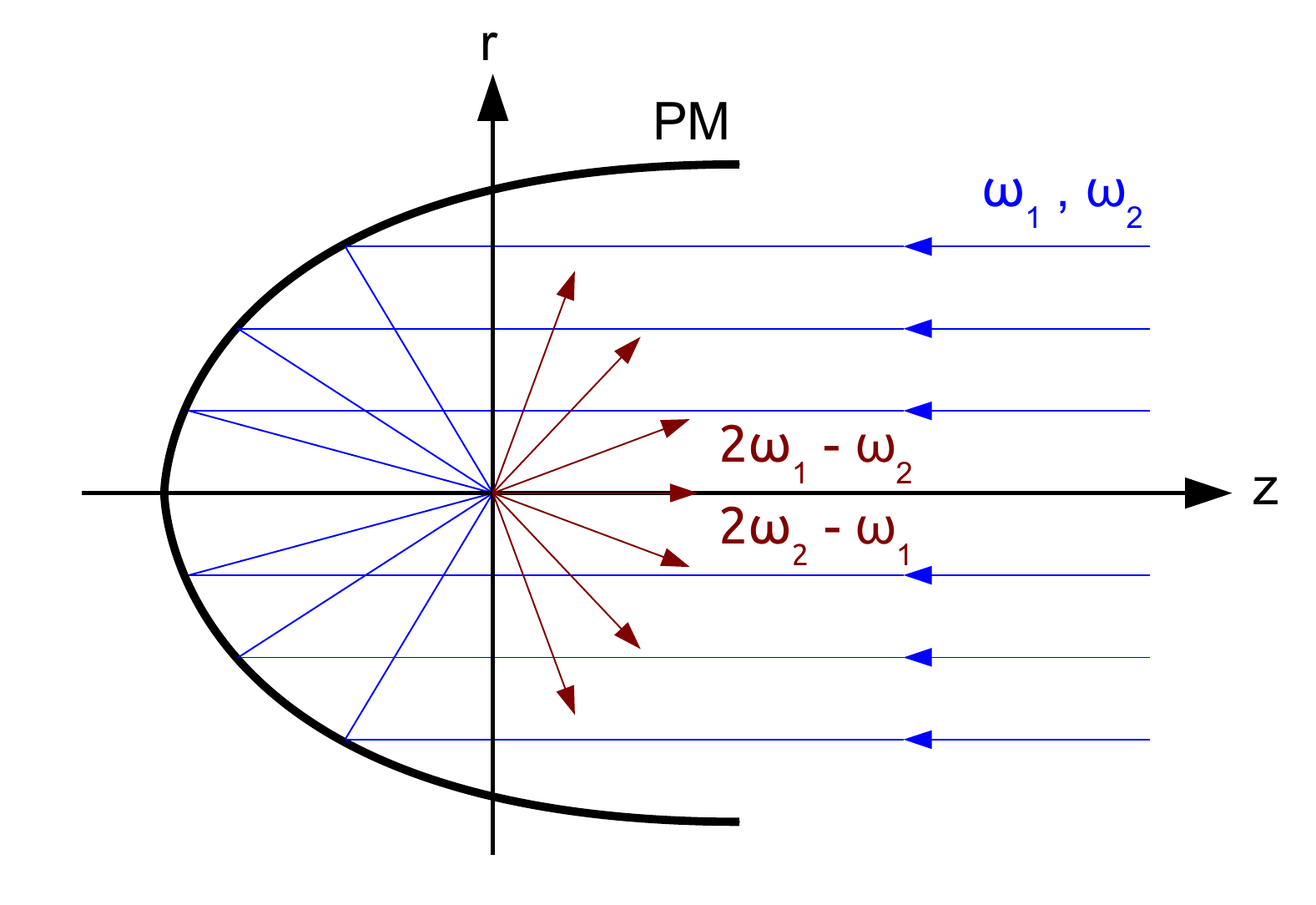}
	\caption{Field configuration: an incident laser beam (blue) with co-propagating frequencies $\omega_1$ and $\omega_2$ is focused by a parabolic mirror (PM). At the focal point, the field strength of the radiation is the highest and photons of frequencies $\omega_a =  2\omega_{1} - \omega_{2}$ and $\omega_b=2\omega_{2}-\omega_{1}$ (red) can be generated. }
	\label{fig:exp_setup}
\end{figure}

The present work proposes a scheme which could allow the experimental study of QED processes (and possibly other processes involving photons in the initial state) in planned high-intensity laser infrastructures. More precisely, the emphasis is on mixing processes due to photon-photon interactions, where two incident laser beams with different frequencies $\omega_{1}$ and $\omega_{2}$ interact with each other to generate harmonics (this study focuses on the generation of $\omega_a = 2\omega_{1} - \omega_{2}$ and $\omega_b=2\omega_{2}-\omega_{1}$). The field configuration considered is depicted in Fig. \ref{fig:exp_setup}: two radially polarized laser beams are initially co-propagating and tightly focused by a parabolic mirror with a High Numerical Aperture (HNA). A similar experimental setup has been used successfully in \cite{payeur2012} to accelerate electrons, but has never been applied to QED study. Mixing processes then occur close to the focal point where the electromagnetic radiation reaches its highest field strength. It will be shown that this geometry can generate a measurable number of emitted photons while circumventing some technical and experimental challenges. It also opens the door for the study of other QED processes such as pair production or Compton scattering.

There have been some proposals to detect these mixing effects using other field configurations. One of the earlier works on this topic considered the four wave mixing process of three linearly polarized plane and Gaussian waves \cite{rozanov1993four,moulin2000,Moulin1999137}. This was tested experimentally using two strong counterpropagating beams and one weaker beam, interacting at an angle \cite{moulin2000}. It was concluded that the intensity level used in the experiment was not high enough to detect the QED effect, although it was possible to put an experimental upper limit on the photon-photon cross-section of $\sigma_{\rm QED } < 1.5 \times 10^{-48}$ cm$^{2}$ \cite{moulin2000}. Similar theoretical investigations have been performed, for other promising field configurations, using the crossing of three Gaussian beams with optimized polarizations \cite{MarklundPRL} and by considering the collision of two short pulses \cite{1367-2630-14-10-103002}. Also, the diffraction of an X-ray probe on a tightly focused beam has been studied \cite{PhysRevLett.97.083603}. Recently, the generation of radiation from a single monochromatic strongly focused beam was considered \cite{PhysRevLett.107.073602}, where the nonlinearly induced field has the same frequency as the incident laser field. In this particular case, it is argued that the two signals can be discriminated by using polarization because the generated radiation has a field component orthogonal to the incident linearly polarized field. Finally, the distorsion of the wavefront from self-induced QED interactions was proposed for Gaussian beams \cite{PhysRevA.90.063803}.

The scheme considered in this article is based on a novel combination of many techniques: one single beam \cite{PhysRevLett.107.073602} with spectral shaping \cite{Lassonde:11}, radial polarization \cite{april2010ultrashort,payeur2012}, frequency mixing \cite{PhysRevLett.111.060404,MarklundPRL} and tight focusing \cite{PhysRevLett.107.073602,payeur2012}. In particular, it is assumed that the incident field has many frequency components that can mix through nonlinear effects, generating a signal at a different frequency. This approach offers an interesting alternative to the above cited techniques because it allows for the utilization of tightly focused radially polarized beams, which enhance QED nonlinear effects \cite{PhysRevLett.111.060404}.

%All of the above cited approaches entail experimental challenges.

%and exploits the fact that the generated signal has a different frequency due to mixing processes.

%It was demonstrated that the radiation is mostly emitted in a region where the reflection of the incident beam is weaker. However, given the weakness of the generated signal, it may be challenging to discriminate the emitted signal of interest from the incident beam reflection as they both have the same frequency. 

In this article, the photon number and space distribution are evaluated for mixing processes occurring at the focus of the HNA parabola. Describing the electromagnetic field theoretically for such a configuration requires a special theoretical treatment. In this work, the Stratton-Chu vector diffraction is introduced to perform this task \cite{PhysRev.56.99,Varga:00}, which, to the best of our knowledge, has never been used to study non-linear effects in vacuum. The rationale behind this choice is that in a tightly focused configuration, the beam divergence angle can obey $\epsilon >1$ rad (for instance, this can be achieved experimentally in a configuration where the numerical aperture is NA$ \gtrsim 0.84$ \cite{PhysRevLett.91.233901}). In this regime, techniques based on the paraxial approximation do not converge \cite{1367-2630-8-8-133}. On the other hand, the Stratton-Chu formalism does not suffer from these limitations.    

The strategy to evaluate the photon distribution is divided in three main steps:
1. Fixing the incident field.
%The incident field has to be similar to the actual laser field used in experiments. Typically, this will be modelled by Gaussian beams, although it is possible to use other fields. 
2. Calculation of the field close to the focal point.
%When the laser is focused tightly by the parabolic mirror, the paraxial approximation does not hold anymore and thus, the Gaussian beam expressions are not accurate. In this work, the determination of the field at the focal region proceeds by using the Stratton-Chu integral \cite{PhysRev.56.99,Varga:00}, which allows to solve the electromagnetic boundary value problem with a minimal set of approximation.
3. Determination of the generated radiation.
%The generated radiation is obtained by solving Maxwell's equations with polarization and magnetization given by the QED effective theory. The solution is obtained by linearizing Maxwell's equations and by using the usual retarded potential formulation. All of these steps will be detailed in the following.
%
Each step will be detailed in the following. QED units in which $\hbar = c  = m = \epsilon_{0} = \mu_{0} = 1$ (where $m$ is the electron mass) and $e=\sqrt{4\pi\alpha}$ are used throughout the calculations. %In this case, the unit length is $l_{\rm u} = \hbar/(mc) \sim 3.86159 \times 10^{-13}$~m (0.386 pm) while the unit time is $t_{\rm u} = \hbar / (mc^{2}) \sim 1.2880885 \times 10^{-21}$~s (1.288 zs), as compared to atomic units: $l_{\rm a.u.} = 0.052$ nm and $t_{\rm a.u.} = 24 \times 10^{-18}$ s (24 as). In these units, the electric field strength is expressed as a fraction of the Schwinger field $E_{S} = 1.3 \times 10^{18}$ V/m.

%\section{Maxwell's equation in vacuum and photon-photon interaction}
%\label{sec:max_eq}

The starting point of this work is the set of time-dependent Maxwell's equations given by
\begin{eqnarray}
\label{eq:maxw1}
\frac{\partial \mathbf{B}(\mathbf{r},t)}{\partial t} &=& - \nabla \times \mathbf{E}(\mathbf{r},t)\;,\; \nabla \cdot \mathbf{B}(\mathbf{r},t) = 0 ,\\
\label{eq:maxw2}
\frac{\partial \mathbf{D}(\mathbf{r},t)}{\partial t} &=& \nabla \times \mathbf{H}(\mathbf{r},t) \;,\; \nabla \cdot \mathbf{D}(\mathbf{r},t) = 0 , %- \nabla \times \mathbf{M}(\mathbf{r},t) - \frac{\partial \mathbf{P}(\mathbf{r},t)}{\partial t} \\
%\label{eq:maxw3}
%\nabla \cdot \mathbf{B}(\mathbf{r},t) &=& 0 ,\\
%\label{eq:maxw4}
%\nabla \cdot \mathbf{D}(\mathbf{r},t) &=& 0 ,
\end{eqnarray}
where $\mathbf{r}$ is the space coordinate, $t$ is the time and $\mathbf{E}$ and $\mathbf{B}$ are the electric and magnetic fields, respectively. 
The displacement field $\mathbf{D}$ and magnetizing field $\mathbf{H}$ are given in terms of the polarization $\mathbf{P}$ and magnetization $\mathbf{M}$ as
\begin{eqnarray}
\label{eq:DH}
\mathbf{D}(\mathbf{r},t) &=& \mathbf{E}(\mathbf{r},t)+ \mathbf{P}(\mathbf{r},t),\\
\mathbf{H}(\mathbf{r},t) &=& \mathbf{B}(\mathbf{r},t) - \mathbf{M}(\mathbf{r},t).
\end{eqnarray}
So far, this corresponds exactly to classical electromagnetic theory in matter where the polarization and magnetization characterize the response of materials to the electromagnetic field. According to the QED effective action (Heisenberg-Euler action \cite{dunne2005heisenberg}), the ``vacuum'' also behaves in a similar way due to photon-photon interactions. Within this approximation, all fermionic degrees of freedom are integrated out using the techniques of low energy effective quantum field theories. In the weak field limit, when $|\mathbf{E}|,|\mathbf{B}| \ll E_{S}$, in the low photon energy limit, when $\hbar \omega \ll mc^{2}$, and for wavelengths smaller than the Compton wavelength \cite{PhysRevD.2.2341}, this action yields the Maxwell's equations described previously, with polarization and magnetization vector fields given by \cite{dunne2005heisenberg}
\begin{eqnarray}
\label{eq:pol}
\mathbf{P}(\mathbf{r},t) &=& a \biggl\{ 2\left[\mathbf{E}^{2}(\mathbf{r},t)- \mathbf{B}^{2}(\mathbf{r},t)   \right] \mathbf{E}(\mathbf{r},t) \nonumber \\
&&+ 7 \left[\mathbf{E}(\mathbf{r},t) \cdot \mathbf{B}(\mathbf{r},t) \right]\mathbf{B}(\mathbf{r},t) \biggr\},\\
\label{eq:mag}
\mathbf{M}(\mathbf{r},t) &=& a \biggl\{ 2\left[\mathbf{B}^{2}(\mathbf{r},t) - \mathbf{E}^{2}(\mathbf{r},t)   \right] \mathbf{B}(\mathbf{r},t) \nonumber \\
&&+ 7 \left[\mathbf{E}(\mathbf{r},t) \cdot \mathbf{B}(\mathbf{r},t) \right] \mathbf{E}(\mathbf{r},t) \biggr\},
\end{eqnarray}
where the constant $a:= \frac{4\alpha^{2}}{45}$. Maxwell's equations, along with the expression of the polarization and magnetization obtained from the QED effective action, govern the time evolution of the electric and magnetic fields with photon self-interaction. These equations (\ref{eq:DH})-(\ref{eq:mag}) can be turned into wave equations for the electric and magnetic fields.
%
%\begin{eqnarray}
%\label{eq:wave_eq_E}
%\left[ \partial_{t}^{2} - \nabla^{2} \right] \mathbf{E}(\mathbf{r},t)  = \nonumber \\
%- \partial_{t}  \left[ \nabla \times \mathbf{M}(\mathbf{r},t) \right] -\partial_{t}^{2} \mathbf{P}(\mathbf{r},t) + \nabla \left[\nabla \cdot \mathbf{P}(\mathbf{r},t)  \right], \\
%\label{eq:wave_eq_B}
%\left[ \partial_{t}^{2} - \nabla^{2} \right] \mathbf{B}(\mathbf{r},t)  = \nonumber \\  \partial_{t}  \left[ \nabla \times \mathbf{P}(\mathbf{r},t) \right] -\nabla^{2} \mathbf{M}(\mathbf{r},t) + \nabla \left[\nabla \cdot \mathbf{M}(\mathbf{r},t)  \right].
%\end{eqnarray}
%
They form a coupled set of six nonlinear partial differential equations, which can be solved by linearization, writing $\mathbf{E}(\mathbf{r},t) = \tilde{\mathbf{E}}(\mathbf{r},t) + \mathbf{E}_{\rm foc}(\mathbf{r},t)$ and $\mathbf{B}(\mathbf{r},t) = \tilde{\mathbf{B}}(\mathbf{r},t) + \mathbf{B}_{\rm foc}(\mathbf{r},t)$,
%
%\begin{eqnarray}
%\mathbf{E}(\mathbf{r},t) &=& \tilde{\mathbf{E}}(\mathbf{r},t) + \mathbf{E}_{\rm foc}(\mathbf{r},t), \\
%\mathbf{B}(\mathbf{r},t) &=& \tilde{\mathbf{B}}(\mathbf{r},t) + \mathbf{B}_{\rm foc}(\mathbf{r},t),
%\end{eqnarray}
%
where $\mathbf{E}_{\rm foc},\mathbf{B}_{\rm foc}$ are the focused external fields and $\tilde{\mathbf{E}},\tilde{\mathbf{B}} \propto \alpha^{2}$ are the weak fields generated by the nonlinear interaction: %The external field propagates freely in the vacuum and thus, obeys the homogeneous wave equation. 
they obey $| \tilde{\mathbf{E}}(\mathbf{r},t)|,| \tilde{\mathbf{B}}(\mathbf{r},t)| \ll  |\mathbf{E}_{\rm foc}(\mathbf{r},t)|,|\mathbf{B}_{\rm foc}(\mathbf{r},t)|$. The wave equations
%, Eqs. (\ref{eq:wave_eq_E})-(\ref{eq:wave_eq_B}), 
can then be solved  using retarded potentials. The latter can be simplified furthermore within the far field approximation \cite{PhysRevA.82.032114}: %assuming the distance $R$, where the emitted field is measured, is much larger than the interaction region $L$ \footnote{For a tightly focused laser field, this is a few wavelengths $L \sim \lambda$}, that is $R \gg L$. Finally, using vector calculus identities, the solution can be simplified to \cite{PhysRevA.82.032114}
\begin{eqnarray}
\label{eq:E_total}
\tilde{\mathbf{E}}_{a}(\mathbf{r},t) &=& \frac{\omega_{a}^{2}e^{-i\omega_{a} (t-R)}}{4\pi R}  \int \biggl[  - \hat{R} \times \mathbf{M}_{a,\rm foc}(\mathbf{s})  + \mathbf{P}_{a,\rm foc}(\mathbf{s}) \nonumber \\
&&- \hat{R} \left[\hat{R} \cdot \mathbf{P}_{a,\rm foc}(\mathbf{s})  \right]\biggr] e^{-i\omega_{a}  \hat{R}\cdot \mathbf{s} } d^{3} \mathbf{s}, \\
\label{eq:B_total}
\tilde{\mathbf{B}}_{a}(\mathbf{r},t) &=& \frac{\omega_{a}^{2}e^{-i\omega_{a} (t-R)}}{4\pi R} \int \biggl[   \hat{R} \times \mathbf{P}_{a,\rm foc}(\mathbf{s})  + \mathbf{M}_{a,\rm foc}(\mathbf{s}) \nonumber \\
&&- \hat{R} \left[\hat{R} \cdot \mathbf{M}_{a,\rm foc}(\mathbf{s})  \right]\biggr] e^{-i\omega_{a}  \hat{R}\cdot \mathbf{s}} d^{3} \mathbf{s},
\end{eqnarray}
for photons generated at frequency $\omega_{a}$, where $R:=|\mathbf{r}|$ and $\hat{R}:=\mathbf{r}/R$. The number of photons at this frequency $N_{a}$ and their spatial distribution on the surface of the detector $\mathcal{S}_{d}$ can be estimated from the time-averaged Poynting vector:
\begin{eqnarray}
\frac{dN_{a}}{d\mathcal{S}_{d}} &=& \frac{\tau}{\omega_{a}} \left|\langle \mathbf{S}_{a} \rangle \right|= \frac{\tau}{2\omega_{a}} \left| \mathrm{Re} \left[  \tilde{\mathbf{E}}_{a} \times \tilde{\mathbf{B}}_{a}^{*} \right]\right|,
\end{eqnarray}
where $\langle \cdots \rangle$ stands for the time average and $\tau$ is the pulse duration, assuming a rectangular pulse time profile. To complete the calculation, the evaluation of $\mathbf{E}_{a,\rm foc}$ and $\mathbf{B}_{a,\rm foc}$ is required. Both correspond to the focused laser field close to the focal region, which is now discussed in the context of the Stratton-Chu diffraction formulation.

%\section{Electromagnetic field at the focus: the Stratton-Chu integral}
%\label{sec:emf_SC}

When a wave is strongly focused by a parabolic mirror with a beam divergence angle $\epsilon > 1$ rad, the paraxial approximation is unreliable and other techniques have to be employed to solve Maxwell's equations \cite{April:10,1367-2630-8-8-133}. The Stratton-Chu integrals give the electromagnetic field generated by an opened emitting surface illuminated by an incoming laser field $\mathbf{E}_{\rm inc},\mathbf{B}_{\rm inc}$,  at frequency $\omega_{a}$, for any perfectly conducting surfaces. They are given explicitly by \cite{PhysRev.56.99,Varga:00}:
%
%For a perfectly conducting surface, the electromagnetic field generated by an opened emitting surface illuminated by an incoming laser field $\mathbf{E}_{\rm inc},\mathbf{B}_{\rm inc}$ at frequency $\omega_{a}$ is given by the following Stratton-Chu integrals \cite{PhysRev.56.99,Varga:00}:
%
%\begin{eqnarray}
%\label{eq:E_SC_cond}
%\mathbf{E}_{a,\rm foc}(\mathbf{r},t) &=& \frac{1}{4\pi} \int_{\mathcal{S}}\bigg\{ 
%ik\mathbf{J}_{\mathcal{S}}G 
%+ \rho_{\mathcal{S}}\nabla_{\mathcal{S}} G
%\biggr\} d\mathcal{S} \nonumber \\
%&&- \frac{1}{4\pi i k} \oint_{\partial \mathcal{S}} (\nabla_{\mathcal{S}} G) (\hat{\mathbf{n}} \times \mathbf{J}_{\mathcal{S}}) \cdot d\boldsymbol{\ell}, \\
%\label{eq:B_SC_cond}
%\mathbf{B}_{a,\rm foc}(\mathbf{r},t) &=& \frac{1}{4\pi} \int_{\mathcal{S}}\bigg\{ 
% \mathbf{J}_{\mathcal{S}}\times \nabla_{\mathcal{S}} G 
%\biggr\} d\mathcal{S} \nonumber \\
%&&- \frac{1}{4\pi i k} \oint_{\partial \mathcal{S}} (\nabla_{\mathcal{S}} G) \rho_{\mathcal{S}} \hat{\mathbf{n}} \cdot d\boldsymbol{\ell}, 
%\end{eqnarray}
%
\begin{eqnarray}
\label{eq:SC_final_E}
\mathbf{E}_{a,\rm foc}(\mathbf{r},t) &=& \frac{1}{2\pi} \int_{\mathcal{S}}\bigg\{ 
ik(\hat{\mathbf{n}} \times \mathbf{B}_{\rm inc})  
+ (\hat{\mathbf{n}} \cdot \mathbf{E}_{\rm inc})\nabla_{\mathcal{S}} 
\biggr\} G d\mathcal{S} \nonumber \\
&&- \frac{1}{2\pi i k} \oint_{\partial \mathcal{S}} (\nabla_{\mathcal{S}} G) \left[ \hat{\mathbf{n}} \times (\hat{\mathbf{n}} \times \mathbf{B}_{\rm inc}) \right] \cdot d\boldsymbol{\ell}, \\
\label{eq:SC_final_B}
\mathbf{B}_{a,\rm foc}(\mathbf{r},t) &=& \frac{1}{2\pi} \int_{\mathcal{S}}\bigg\{ 
 (\hat{\mathbf{n}} \times \mathbf{B}_{\rm inc})\times \nabla_{\mathcal{S}} G 
\biggr\} d\mathcal{S} \nonumber \\
&&- \frac{1}{2\pi i k} \oint_{\partial \mathcal{S}} (\nabla_{\mathcal{S}} G)  (\hat{\mathbf{n}}\cdot \mathbf{E}_{\rm inc})  \hat{\mathbf{n}}\cdot d\boldsymbol{\ell}, 
\end{eqnarray}
where $\mathcal{S}$ is the surface of the mirror, $\boldsymbol{\ell}$ is the tangent vector on the mirror opening $\partial \mathcal{S}$, $\hat{\mathbf{n}}$ is the unit vector normal to the mirror, $k$ is the wave vector and $G$ is the Green's function. The parabolic shape of the mirror is accounted for in the surface integral by setting $r^{2}_{\mathcal{S}} = 4f(z_{\mathcal{S}}+f)$ for the coordinates on the parabola (in cylindrical coordinates $z$ and $r$), where $f$ is the focal length. %It is also assumed that the charge density and current are induced by the incident and reflected field \footnote{In principle, the incoming beam would be diffracted by the mirror opening. We neglect this phenomenon in this work as the beam wavelength is much smaller than the mirror radius.}.
%%
%\begin{eqnarray}
%\rho_{\mathcal{S}} &=& 2\hat{\mathbf{n}} \cdot \mathbf{E}_{\rm inc}, \\
%\mathbf{J}_{\mathcal{S}} &=& 2\hat{\mathbf{n}} \times \mathbf{B}_{\rm inc},
%\end{eqnarray}
%%
%where $\mathbf{E}_{\rm inc},\mathbf{B}_{\rm inc}$ are the incoming laser electric and magnetic fields, respectively. The factor of 2 on the RHS is included to take the reflected beam into account \cite{Varga:00}. Finally, the Stratton-Chu integrals become 
%
%\begin{eqnarray}
%\label{eq:SC_final_E}
%\mathbf{E}_{a,\rm foc}(\mathbf{r},t) &=& \frac{1}{2\pi} \int_{\mathcal{S}}\bigg\{ 
%ik(\hat{\mathbf{n}} \times \mathbf{B}_{\rm inc})  
%+ (\hat{\mathbf{n}} \cdot \mathbf{E}_{\rm inc})\nabla_{\mathcal{S}} 
%\biggr\} G d\mathcal{S} \nonumber \\
%&&- \frac{1}{2\pi i k} \oint_{\partial \mathcal{S}} (\nabla_{\mathcal{S}} G) \left[ \hat{\mathbf{n}} \times (\hat{\mathbf{n}} \times \mathbf{B}_{\rm inc}) \right] \cdot d\boldsymbol{\ell}, \\
%%
%\label{eq:SC_final_B}
%\mathbf{B}_{a,\rm foc}(\mathbf{r},t) &=& \frac{1}{2\pi} \int_{\mathcal{S}}\bigg\{ 
% (\hat{\mathbf{n}} \times \mathbf{B}_{\rm inc})\times \nabla_{\mathcal{S}} G 
%\biggr\} d\mathcal{S} \nonumber \\
%&&- \frac{1}{2\pi i k} \oint_{\partial \mathcal{S}} (\nabla_{\mathcal{S}} G)  (\hat{\mathbf{n}}\cdot \mathbf{E}_{\rm inc})  \hat{\mathbf{n}}\cdot d\boldsymbol{\ell}. 
%\end{eqnarray}
%
It is clear from Eqs.  (\ref{eq:SC_final_E})-(\ref{eq:SC_final_B}) that any type of incoming laser field can be used to evaluate the field at the focus, giving us the flexibility to study various field configurations. Moreover, the Stratton-Chu equations are integral solutions to Maxwell's equations and thus, should describe accurately any kind of fields in a tightly focused configuration. This is an advantage to achieve a realistic description of an experimental setup using an HNA parabola, as compared to other analytical methods based on the paraxial approximation \cite{1367-2630-8-8-133} or an infinite parabola \cite{April:10}.

The incident laser field considered here is a collimated radially polarized Gaussian laser beam propagating in the $-\hat{z}$ direction. Because it is collimated, the beam divergence is $\epsilon_{\rm inc} \ll 1$ and the paraxial approximation can be employed (as opposed to $\mathbf{E}_{\rm foc},\mathbf{B}_{\rm foc}$). The use of the paraxial approximation in the incoming beam induces a negligible error $O(\epsilon_{\rm inc})$ in the calculation of the field at the focus. The expression of the laser field is then given by \cite{1367-2630-8-8-133}:
\begin{eqnarray}
\label{eq:radial_Er}
E_{\mathrm{inc},r}(\mathbf{r}_{\mathcal{S}},t) &=&- E_{0} \frac{2r_{\mathcal{S}}}{k w_{0}^{2}} e^{- \frac{r^{2}_{\mathcal{S}}}{w_{0}^{2}} -i\omega t - i k z_{\mathcal{S}}} , \\
\label{eq:radial_Ez}
E_{\mathrm{inc},z}(\mathbf{r}_{\mathcal{S}},t) &=& i E_{0} \frac{4}{k^{2} w_{0}^{2}} \left[ 1- \frac{r^{2}_{\mathcal{S}}}{w_{0}^{2}} \right] e^{- \frac{r^{2}_{\mathcal{S}}}{w_{0}^{2}} -i\omega t - i k z_{\mathcal{S}}} , \\
\label{eq:radial_Bth}
B_{\mathrm{inc},\theta}(\mathbf{r}_{\mathcal{S}},t) &=& E_{0} \frac{2r_{\mathcal{S}}}{k w_{0}^{2}} e^{- \frac{r^{2}_{\mathcal{S}}}{w_{0}^{2}} -i\omega t - i k z_{\mathcal{S}}},
\end{eqnarray}
where $w_{0}$ is the beam width, $k$ is the wave number and $E_{0}$ is a normalization constant fixed from the time-average value of the Poynting vector, which is given by $E_{0} = k \sqrt{\frac{2U}{\pi \tau}}$, where $U$ is the energy per pulse. This choice of incident field is justified by the fact that radially polarized beams can be focused on a smaller region compared to linear polarization, leading to a higher field strength \cite{PhysRevLett.91.233901}. Moreover, they maximize the vacuum polarization and magnetization close to the focal spot because their only nonzero field component is longitudinal ($E_{z} \neq 0$), which is a major advantage over other polarizations. Finally, they can be obtained from linearly polarized beams using the technique described in \cite{leuchs_2005,payeur2012}. They have been generated at high power ($\approx$ 200 TW) and the extension of this technique to higher power should be feasible, in principle \cite{kieffer}. At such high power, nonlinear effects in glass, affecting the beam quality, appear when the fluence becomes too high. However, this can be controlled by enlarging the beam if required.

All the ingredients to compute the photon distribution have been discussed. To summarize, the first step is the calculation of the field at the focus using Eqs. \eqref{eq:SC_final_E} and \eqref{eq:SC_final_B}, for two frequencies $\omega_{1}$ and $\omega_{2}$. This is performed numerically with Gauss-Legendre quadrature. In the second step, the components of $\mathbf{P}_{\rm foc}$ and $\mathbf{M}_{\rm foc}$ with frequencies $\omega_{a} = 2\omega_{1}-\omega_{2}$ and  $\omega_{b} = 2\omega_{2}-\omega_{1}$ are extracted analytically and evaluated numerically. Finally, the generated field is computed numerically using Eqs. \eqref{eq:E_total} and \eqref{eq:B_total}. 

In numerical calculations, the wavelengths of the incoming laser field are set to $\lambda_{1} = 820$ nm and $\lambda_{2} = 780$ nm, which can be obtained from a 800 nm beam by a spectral pulse shaping technique \cite{Lassonde:11}. Consequently, the radiation from mixing processes in the vacuum will be emitted at $\lambda_a \approx 864$ nm and $\lambda_b \approx 744$ nm. A broad spectrum for the incident beams, required to describe short pulses, can also be considered in principle. The effect of this will be studied in future work. 

%This would be possible  require some more work to design an experimental implementation. 
 
\begin{figure}[h]
	\centering
		\includegraphics[width=0.40\textwidth]{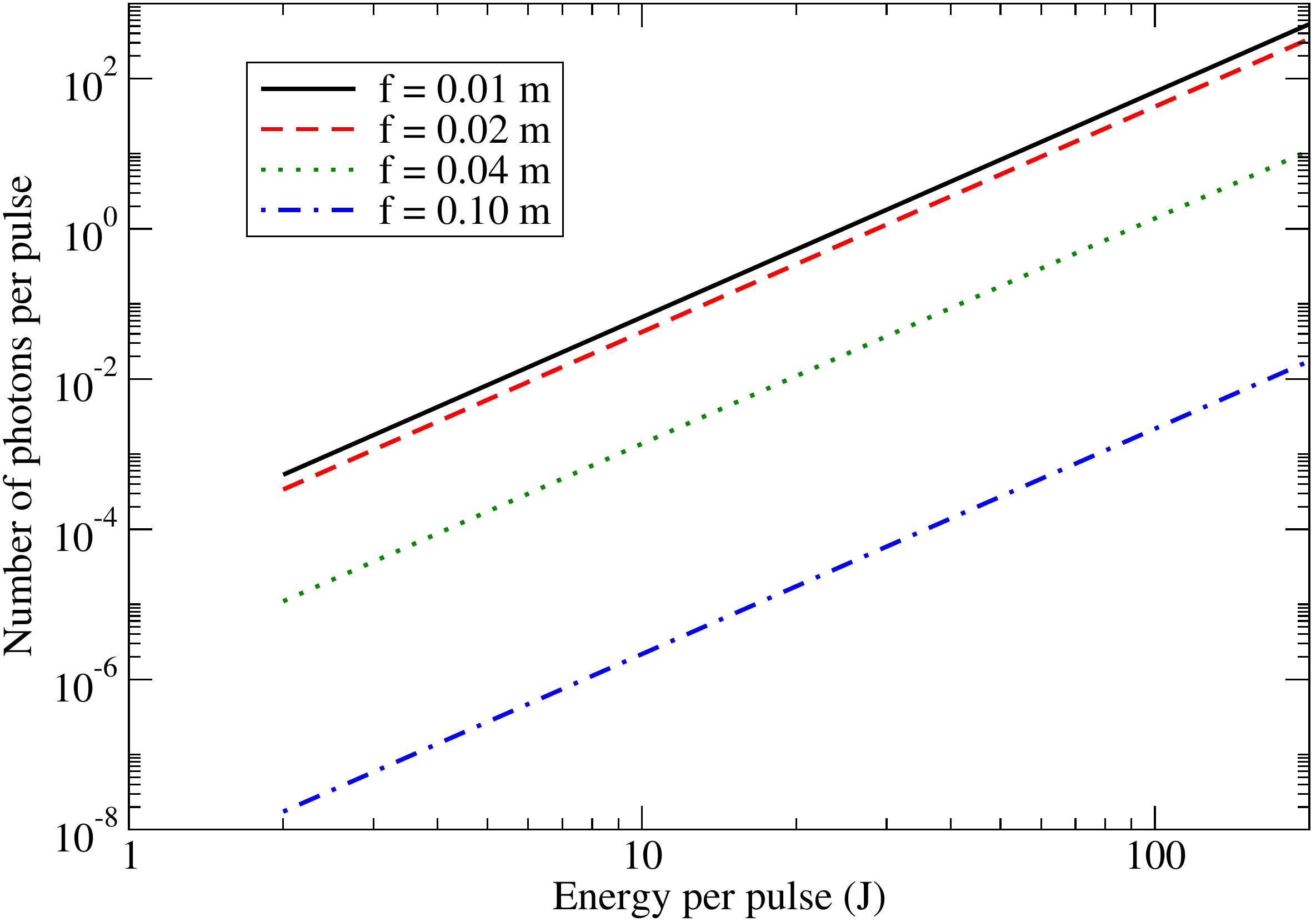}
	\caption{Total number of photons emitted at $\omega_a = 2\omega_{1}-\omega_{2}$, as a function of the energy per pulse for focal length $f=0.01, 0.02, 0.04$ and $0.10$ m. The size of the parabola aperture is $r_{\rm max} = 0.08$ m. Incident pulse: $\lambda_{1} = 820$ nm and $\lambda_{2} = 780$ nm with a Gaussian profile $w_0 = 0.03$ m and pulse duration $\tau=30$ fs. }
	\label{fig:nu_photons_focal}
\end{figure}

Integrated over all angles of emission, the total number of photons $N$ is sensitive to the incident laser pulse characteristics (energy and pulse duration) as well as the parabola parameters (i.e. focusing parameters). The total number of photons $N$ emitted at $\omega_a$ is shown in Fig. \ref{fig:nu_photons_focal}, as a function of the energy per pulse, for different focal lengths. Clearly, for smaller focal lengths, the number of photons can be enhanced by many orders of magnitude due to the larger focused field. The trend is similar to the results found in \cite{PhysRevLett.107.073602}. The scaling of $N$ with the pulse energy ($U$) shown in  figure \ref{fig:nu_photons_focal} can be obtained analytically from the scaling of the field and the expressions for the emitted radiation: the number of photons scales like $N \propto U^{3}/\tau^{2}$, which explains the rapid rise of the photon production rate with the energy per pulse. 

\begin{figure}
	\centering
		\includegraphics[width=0.40\textwidth]{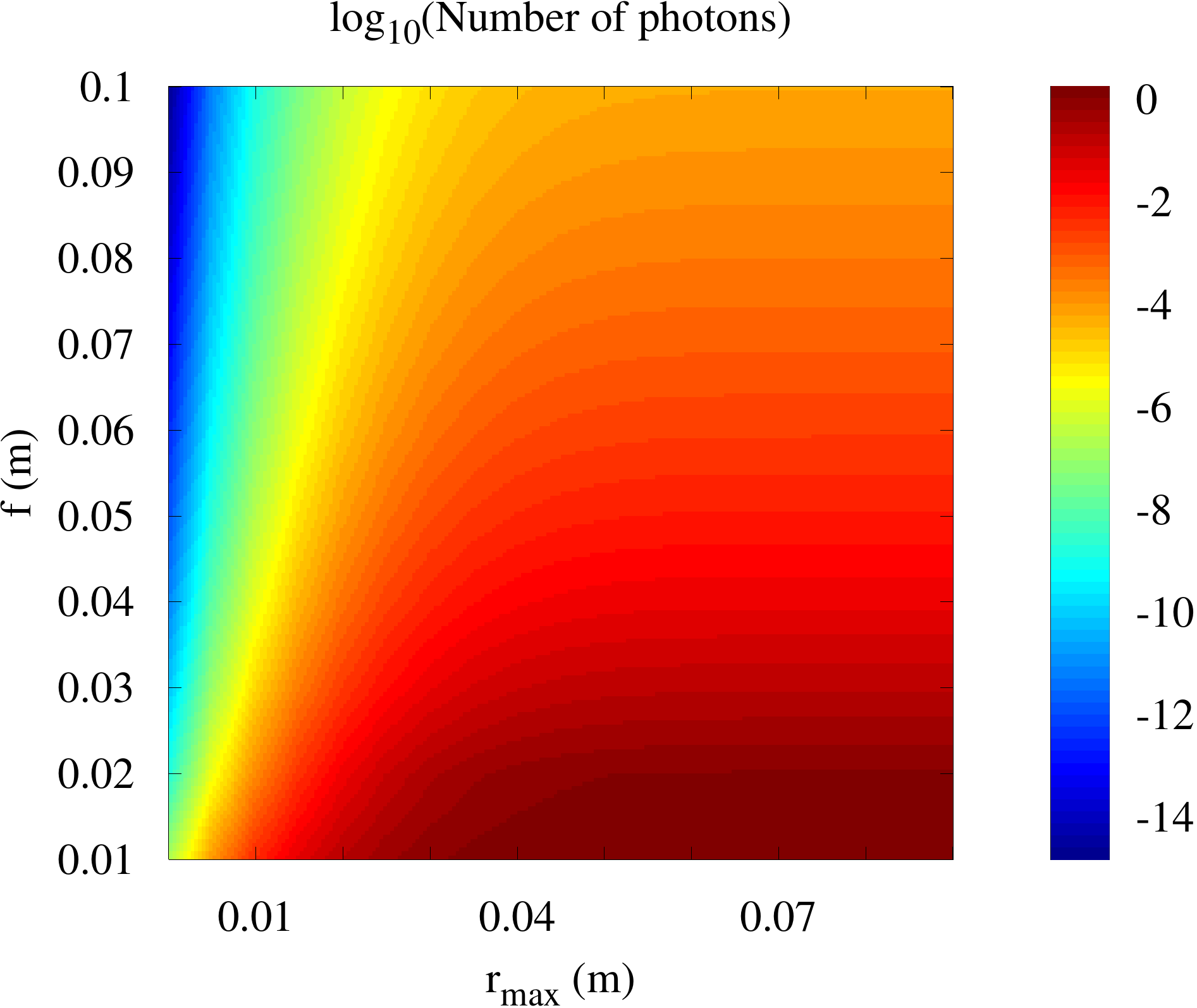}
	\caption{Total number of photons emitted at $\omega_a = 2\omega_{1}-\omega_{2}$, as a function of the focal length $f$ and the parabola aperture $r_{\rm max}$.  Incident pulse: $\lambda_{1} = 820$ nm and $\lambda_{2} = 780$ nm with a Gaussian profile $\omega_0 = 0.03$ m, pulse duration $\tau=30$ fs and energy $U=30$ J. }
	\label{fig:photon_number_f_rmax}
\end{figure}

In the following numerical calculations, we consider a 1 PW laser, which is now available \cite{RevModPhys.84.1177}, with a pulse length of $\tau = 30$ fs and an energy per pulse of $U=30$ J.
The total number of photons emitted at $\omega_a$ as a function of focal length ($f$) and aperture size ($r_{\rm max}$) is shown  in Fig. \ref{fig:photon_number_f_rmax}. This result demonstrates that the number of photons emitted saturates at a certain value of $r_{\rm max}$ where the non-reflected tail of the Gaussian beam is negligible. More interesting however is the non-linear increase of the number of emitted photons as the focal length is decreased or as the aperture size is increased: this effect is caused by the higher field strength attained as $f$ becomes smaller and $r_{\rm max}$ becomes larger. This can be used to enhance the signal from wave mixing in vacuum. 

\begin{figure}
	\centering
		\includegraphics[width=0.40\textwidth]{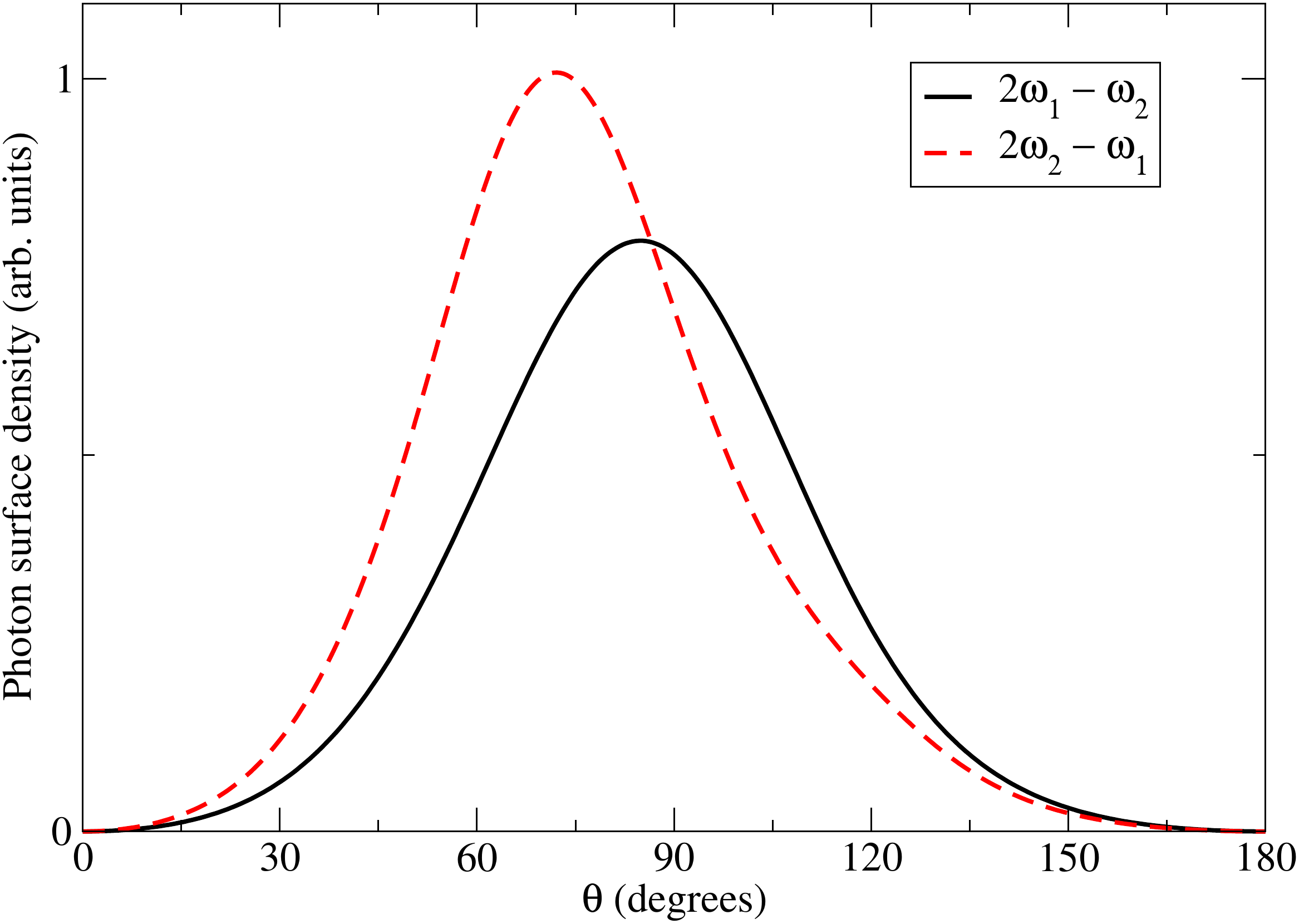}
	\caption{Photon density as a function of the angle of emission $\theta$ with respect to the optical axis of the parabolic mirror of aperture $r_{\rm max} = 0.08$ m and focal length $f=0.02$ m. Incident pulse: $\lambda_{1} = 820$ nm and $\lambda_{2} = 780$ nm with a Gaussian profile $\omega_0 = 0.03$ m, pulse duration $\tau=30$ fs and energy $U=30$ J. }
	\label{fig:photon_density}
\end{figure}

In Fig. \ref{fig:photon_density}, the photon distribution as a function the angle $\theta$ from the optical axis of the parabola (the $z$-axis) is presented with some optimized parameters. In both mixing cases ($\lambda_{a,b}$), most of the photons are emitted in the range $[40^{\circ},90^{\circ}]$. This should guide future experiments for optimizing the detection system.

The numerical results generally show that in a certain parameter range, when the energy per pulse is large, the focal length is small and when the aperture size is large, a noticeable number of photons can be emitted by vacuum mixing processes. For instance, it is claimed that the planned Apollon high-intensity infrastructure should deliver $\tau \sim$~15 fs pulses at 10 PW, for an energy per pulse of 150 J \cite{apollon}. Using a beam width of $w_{0} = 0.2$ m, $f = 0.02$ m and $r_{\rm max} = 0.15$ m, it would be possible to produce $N \approx 6.5$ photons per shot in our tightly focused geometry. 
%To the best of our knowledge, this is one of the highest predicted photon yield using conditions close to experiments. 
Radially polarized beams are instrumental to reach this result because these beams, close to the focal spot, have a longitudinal field with $B_{\theta} = E_{r}=0, E_{z} \neq 0$ and thus, maximize the value of the Lorentz invariant appearing in the vacuum polarization and magnetization ($\mathbf{P}_{\rm foc}$ and $\mathbf{M}_{\rm foc}$). %This is similar to $e$-dipole pulses \cite{PhysRevLett.111.060404}. 

The main advantage of this geometry however pertains to the alignment and synchronization of the beams: in most scenarios (see \cite{Moulin1999137,MarklundPRL} for instance), one considers the crossing of many laser beams. It is a very challenging technical task to align and synchronize many highly focused counter-propagating short pulse laser beams while preserving a high intensity at the interaction region. This is because in order to obtain the intensity required to observe wave mixing in vacuum with lasers of 1 - 10 PW, the laser has to be focused on a very small focal spot, the size of a wavelength ($\lambda_{\rm focal \; spot} \sim \lambda$). The technique presented in this article circumvents these complications because a single incident beam can be used. Then, the mixing occurs at the focus between its different frequency components and this process can be optimized by using a spectral pulse shaping technique \cite{Lassonde:11}. An experimental challenge remains in producing an incident wavelength pair such that the generated frequencies are outside of the laser background. This could be implemented experimentally by manipulating the frequency of the incident beam (spectral shaping, frequency doubling, etc). Some of these possible other configurations have been studied theoretically (for instance, with $\lambda_{1} = 800$ nm,  $\lambda_{2} = 400$ nm and $\lambda_{a} = 266$ nm) and gave similar results to the ones presented in this article \cite{FFG01}. Other background radiation coming from competing processes can also potentially hide the signal. However, it has been argued that they are orders of magnitude below the QED signal for low enough pressure \cite{MarklundPRA}.

%A more detailed analysis of this geometry, including the effect of short pulse spectra, will be presented in future work.   

%for $I > 10^{20}$ W/cm$^{2}$, the only known method is by focusing a short pulse ($\tau \sim$30 fs) using an off axis paraboloid \cite{Bahk:04}.

In conclusion, the QED photon-photon interactions have been studied theoretically. A scheme, using a tightly focused field configuration, radially polarized beams and spectral shaping of the incident beam, has been proposed. The numerical results obtained demonstrate that the latter could be used to generate a noteworthy number of photons by photon-photon interactions for lasers in the range of 1-10 PW and above, which is accessible with soon available technologies. To perform the calculation, the Stratton-Chu formulation is introduced. This methodology relates the number of photons emitted to experimental parameters ($f,r_{\rm max},w_{0},U,\tau$) and thus, is important for future theoretical investigations in high-intensity laser physics. Finally, the experimental setup suggested could be relevant for the investigation of other physical processes.% involving photons and electrons in the initial state.  

\begin{acknowledgments}
The authors would like to thank Drs. S. Fourmaux, K. Otani, S. Payeur and Prof. J.-C. Kieffer for many valuable discussions on this subject. We also thank RQCHP and Compute Canada for access to massively parallel computer clusters.
\end{acknowledgments}

%\References
%\section*{References}
%\bibliographystyle{iopart-num}
%\bibliographystyle{plain}
\bibliographystyle{apsrev4-1}
\bibliography{bibliography}

\end{document}